\documentclass[aip,jap,preprint]{revtex4-1}

\usepackage{graphicx}
\usepackage{amsmath}
\usepackage{amssymb}
\usepackage{natbib}
\usepackage{bm}
\usepackage{color}
\usepackage{siunitx}

\begin{document}
\title{Microbubble Oscillation on Localized Heat Source Affected by Dissolved Gases in Water}

\author{Nao Hiroshige}
 \thanks{Nao Hiroshige has retired from Mitsubishi Electric Corporation.}
\affiliation{Mitsubishi Electric Corporation, Advanced Technology R\&D Center, 8-1-1, Tsukaguchi-Honmachi, Amagasaki city, Hyogo, 661-8661, Japan.}

 \author{Shunsuke Okai}  
 \author{Xuanwei Zhang}
   \affiliation{Department of Micro Engineering,
Kyoto University,
Kyoto Daigaku-Katsura, Nishikyo-ku, Kyoto 615-8540, Japan.}
 
 \author{Samir Kumar}
  \affiliation{Department of Electronics and Information Engineering,
 Korea University, Sejong 30019, Republic of Korea.}

 \author{Kyoko Namura}
 \email{namura.kyoko.2s@kyoto-u.ac.jp}
\affiliation{Department of Micro Engineering,
Kyoto University,
Kyoto Daigaku-Katsura, Nishikyo-ku, Kyoto 615-8540, Japan.}
 
 \author{Motofumi Suzuki}
 \affiliation{Department of Micro Engineering,
Kyoto University,
Kyoto Daigaku-Katsura, Nishikyo-ku, Kyoto 615-8540, Japan.}

\begin{abstract}
Recently, we demonstrated that the local heating of degassed water can generate water vapor microbubbles and induce a rapid flow around the bubble.
 Although flow generation involves the self-excited oscillation of bubbles at a local heating point, the conditions under which the bubbles oscillate are not fully understood.
 In this study, the dependence of microbubble size and oscillation frequency on the concentration of non-condensable gas in water was investigated.
 A continuous-wave laser beam was focused on a $\beta$-FeSi$_2$ thin film, and water was locally heated using the photothermal conversion properties of the film. 
 The results showed that the lower the concentration of non-condensable gas dissolved in water, the smaller the bubble size and the higher the oscillation frequency. 
Furthermore, it was found that the bubbles oscillate when the amount of non-condensable gas absorbed by the bubbles, i.e., the bubble size, falls below a specific level.
 This study can provide a new understanding of the bubble oscillation mechanism and lead to the development of fluid control technology using bubbles.
\end{abstract}

\noindent
\textcolor{red}{{\small This article has been submitted to Journal of Applied Physics.
https://pubs.aip.org/aip/jap}}


\maketitle

\section{INTRODUCTION}
With the rapid development of integration technology in the semiconductor industry, electronic devices such as high-power devices need to be increasingly smaller. The power density per semiconductor chip of these devices continues to increase, reaching hundreds of watts per square centimeter. \cite{Jungwan2015,Laloya2016} The semiconductor chip, which is confined in an extremely small area, causes the fundamental problem of heat dissipation. The large heat flux generated by the semiconductor chip and insufficient heat dissipation in the device package can increase the temperature of the chip to hundreds of degrees Celsius. This leads to undesirable limitations and the deterioration of device performance.

Water-cooling systems, which use external pumps to circulate water, are conventionally used to cool these devices. 
To improve the efficiency and miniaturization of such water-cooling technologies, flow channels have been miniaturized to the micrometer scale. \cite{Missaggia1989,Dang2010,vanErp2020}
Miniaturization enables the fine channels to increase the surface area for heat exchange.
In the 1980s, it was reported that a water flow of 8.6 cm$^3$/s in a microscale channel using an external pump was capable of dissipating as much as 790 W/cm$^2$. \cite{Tuckerman1981} 
More recently, it has been shown that cooling efficiencies in excess of 1 kW/cm$^2$ can be achieved with a smaller pumping power by creating finer cooling structures closer to the heat-generating parts. \cite{vanErp2020}
Thus, micro-liquid cooling technology contributes to achieving high cooling performance. However, an external pump was used to drive the fluid.
In microfluidic channels, high pressure is required to pump the liquid in the channel owing to the viscous effect of the fluid. 
Using a powerful and bulky pump to drive the coolant fluid adversely affects device miniaturization.
Because the miniaturization of pumps is limited, new technologies for driving liquids in microscale channels without relying on external pumps are required.

The technology used to control fluids in micrometer-scale channels is called microfluidics and is being studied in a wide range of fields.
For example, microfluidics has already been utilized in the chemical industry as a microreactor \cite{Mason2007,Shi2019}, in the biomedical industry as a drug discovery system \cite{Dittrich2006}, or for individual cell and DNA analysis by driving liquids \cite{Abdelmoez2018} within a microscale channel. 
Various fluid pumping technologies have been reported to be suitable for these applications. \cite{Laser2004,Wang2018}
Cooling technology is most efficient if waste heat can be used to drive the fluid.
A promising microfluidic pumping technology that uses heat is a combination of bubbles and heat.
Bubbles can drive fluids through dynamic volume changes \cite{Geng2001,Dijkink2008} and shear forces owing to surface tension gradients, that is, Marangoni forces. \cite{Sammarco1999}
The fluid can be manipulated in situ by placing bubbles at appropriate locations in the microchannel to provide an appropriate temperature distribution.
It has been reported that the flow created by bubbles can be used to collect or deposit small particles and specimens, \cite{Nishimura2014,Lin2016,Fujii2017,Tokonami2020,Dong2023}
manipulate the particle motion, \cite{Taylor2004,Zhao2014,Namura2015,Namura2016,Namura2016JNP,Dai2019,Dara2023}
manipulate the bubble itself, \cite{Zeng2021,Hu2022}
mix fluids, \cite{Liu2002,Taylor2004,Hellman2007,Jones2020}
and much more.
However, it is difficult to achieve a sufficiently strong flow to achieve cooling.

Recently, we discovered a method for driving fluids strongly using water-vapor microbubbles. \cite{Namura2017,Namura2019,Namura2022}
When a laser beam is focused on a thin film of gold nanoparticles, the film absorbs light and converts it into heat, allowing the irradiated area to be used as a local heat source.
When degassed water was locally heated using this heat source, bubbles $\sim$10 \si{\um} in diameter, mainly composed of water vapor, were generated.
The bubbles are exposed to a steep temperature gradient and generate a strong flow in the water.
The flow velocity even reached 1 m/s in the very vicinity of the bubble.
 This flow speed is much higher than that achieved in a conventional microfluidic channel, which has a range of 0.1--10 mm/s. 
This rapid flow near the solid surface is expected to break the boundary layer that sits on the solid-liquid surface and acts as a barrier to heat dissipation, and significantly improving the heat dissipation efficiency.
However, the contribution of the gas composition inside the bubble to the enhanced flow velocity around the bubble is not fully understood.
In addition, the water vapor-rich microbubbles generated in degassed water oscillated at several hundred kilohertz, whereas an air microbubble generated in non-degassed water showed no oscillation. \cite{Namura2020} 
Ohl and his colleagues have also reported similar vapor-rich bubbles and strong convection generation.\cite{Li2017,Nguyen2018,Nguyen2019_2,Nguyen2019} 
 These bubbles have also been found to oscillate at several hundred kilohertz.
Therefore, the oscillation of the bubbles was likely related to the generation of a rapid flow around the microbubbles. 
Other groups have extensively studied the influence of dissolved gases in water on the behavior of locally heated bubbles.
\cite{Deguchi2013,Baffou2014,WangY2017,WangY2018,LiXiaolai2019,Zaytsev2020}
In particular, there are substantial reports on the relatively large vapor bubble explosions that occur at the moment of bubble formation\cite{WangY2017,WangY2018,LiXiaolai2019} and on the continuous growth\cite{Baffou2014,LiXiaolai2019} and shrinkage of bubbles.\cite{Baffou2014,Zaytsev2020} 
However, the effect on sub-MHz order oscillations of bubbles generated above the local heating point is still not fully understood.
To understand how rapidly the flow is generated around a water vapor microbubble, it is important to investigate the relationship between degassing water and bubble oscillation.

In this study, we investigated the dependence of the size and oscillation frequency of a microbubble on the non-condensable gas concentration dissolved in water using the photothermal properties of a $\beta$-FeSi$_2$ thin film. We demonstrated that a lower concentration of non-condensable gas dissolved in water leads to a smaller bubble size and a higher bubble oscillation frequency. 
This is because the amount of non-condensable gas that the bubble takes in changes.
Results also showed that when the bubble absorbed a certain amount of non-condensable gas and the equilibrium size of the bubble became larger than that of the heated region, the bubble stopped oscillating.
This indicates that creating conditions in which direct heating of the gas-liquid interface is possible is important for inducing bubble oscillation.

\section{EXPERIMENTS}
A $\beta$-FeSi$_2$ thin film was prepared as a photothermal conversion film by radio-frequency magnetron sputtering. First, the glass substrate was sonicated in acetone and then placed in a UV ozone cleaner (UV253S, Filgen) for 30 min to improve surface wettability. For $\beta$-FeSi$_2$ film sputtering, a FeSi$_2$ target with a purity of 99.9 \% was used. The base pressure of the sputtering chamber was 2.0$\times$10$^{-4}$ Pa. Then, Ar gas was used as the sputtering gas at a flow rate of 9 sccm. The pressure during the sputtering was fixed at 0.86 Pa by controlling the evacuation rate. The temperature of the glass substrate was fixed at 723 K using a radiation heater. The glass substrate was continuously rotated at 4 rpm during sputtering to obtain a uniform film thickness. Finally, we obtained a 50 nm-thick $\beta$-FeSi$_2$ film on the glass substrate. The optical reflectance and transmittance of the prepared thin films were measured using a single-beam spectrophotometer. The optical absorption of the thin films was calculated by subtracting the optical reflectance and transmittance from unity. The optical absorption was 0.58 at a wavelength of 638 nm, at which the photothermal conversion experiments were conducted.

The prepared $\beta$-FeSi$_2$ film was cut into small pieces, one of which was fixed to the internal wall of a glass cell (10$\times$10$\times$58 mm, F15-G-10, GL Science) using a pair of magnets. The cell was then filled with water at a controlled concentration of non-condensable gas. The preparation of water with various concentrations of non-condensable gas is described below. First, 50.0 mL of the ultrapure water (18.2 M\si{\ohm} cm from Millipore-Direct Q UV3, Merck) in a vacuum vessel was sonicated under vacuum ($\sim$3 kPa at 25 \si{\degreeCelsius}) for 20 min. Next, the pressure was adjusted by introducing O$_2$ gas into the vessel. Finally, water was stirred using a magnetic stirrer for 10 min to equilibrate the O$_2$ gas and water. We obtained water with different non-condensable gas concentrations by varying the pressure in a vacuum vessel. The cell filled with the prepared water was sealed to prevent the diffusion of gases from the air into the water. The concentrations of the O$_2$ and N$_2$ gases in the water in the cells were measured using a gas chromatography system (GC-2010, Shimadzu) after the microbubble generation experiment, as described below. The measured gas concentrations are listed in Table \ref{tb1}.
   \begin{table}[tb]
\centering
\caption{Non-condensable gas concentration of prepared water.}
\begin{tabular}{p{0.8in}p{0.8in}p{0.8in}p{0.8in}} \hline
 & \multicolumn{3}{c}{concentration (mg/L)} \\ 
 ID&O$_2$& N$_2$ & O$_2$+N$_2$ \\ \hline \hline
water 1&1.4 & 2.8 & 4.2  \\\hline
water 2&3.7& 6.0 & 9.7  \\\hline
water 3&4.3& 6.5 & 10.8  \\\hline
water 4&6.6 & 4.9 & 11.5  \\ \hline
\end{tabular}
\label{tb1}
\end{table}
The non-condensable gas concentration (O$_2$ + N$_2$) was successfully varied from 4.2 to 11.5 mg/L.

\begin{figure*}[tbp]
\centerline{\includegraphics[bb = 0 0 533 336, width=15cm]{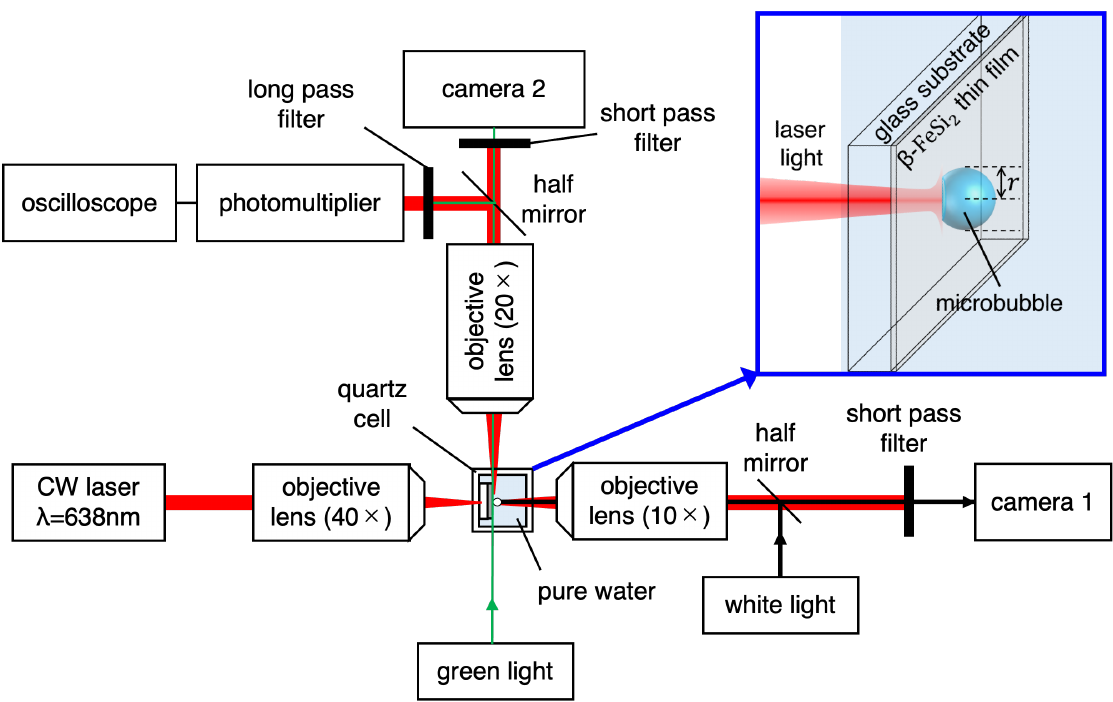}}
\caption{Schematics of the experimental setup. A continuous-wave laser with a wavelength of 638 nm was irradiated on a $\beta$-FeSi$_2$ thin film on a glass substrate from the rear side of the substrate through a 40$\times$ objective lens. The generated bubble was observed from a direction vertical to the substrate surface by a camera through a 10$\times$ objective lens under white light illumination and in a direction parallel to the substrate surface by camera 2 through a 20$\times$ objective lens under green light illumination. A short-pass filter was placed in front of the camera to block the laser light. Laser light scattered from the bubble was guided into the photomultiplier tube to detect the bubble oscillation. The green light was blocked by a long-pass filter placed in front of the photomultiplier tube.}
 \label{fig1}
\end{figure*}

The prepared fluidic cell was placed in an optical setup to conduct experiments on the photothermal generation of microbubbles. Fig.\ref{fig1} shows the optical setup, which enables us to generate microbubbles and measure their size and the oscillation frequency. 
A continuous-wave (CW) laser with a wavelength of 638 nm was focused on the $\beta$-FeSi$_2$ film to photothermally generate a bubble using an objective lens (40$\times$, NA = 0.60). The laser spot size was measured using camera 1 (HXC20, Baumer) with an objective lens (10$\times$, NA = 0.26). For all the bubble generation experiments, the laser spot radius (1/$e^2$) was tuned to 2.75 $\pm$ 0.25 \si{\um}, and the laser power was set to 30 mW. The generated bubbles were observed using camera 2 (FASTCAM Mini AX100, Photron) through an objective lens (20$\times$, NA = 0.40) under green light illumination. 
The frame rate was set to 4000 fps. The radius of the generated bubble, $r$, was measured from the image captured by camera 2, where the spatial resolution was 1 \si{\um}/pixel. Short-pass filters can be inserted in front of cameras 1 and 2 to block the laser.
To detect the bubble oscillations, the laser light scattered from the bubble was guided into a photomultiplier tube (R928, Hamamatsu Photonics), and its signal was obtained using an oscilloscope. The green light used for the observation of the bubble was blocked using a long-pass filter in front of the photomultiplier tube. 
The direction of the laser light scattered from the bubble depends on the bubble size. Therefore, the intensity of the scattered laser light captured by the photomultiplier tube changes periodically along with the periodic oscillation of the bubble.\cite{Namura2020} The oscillation frequency of the bubble was obtained by analyzing the signal using fast Fourier transform (FFT). Simultaneous measurements of the radius and oscillation frequency were repeatedly performed on the bubbles 15, 20, 18, and 18 times in water 1, 2, 3, and 4, respectively.

\section{RESULTS AND DISCUSSION}
\begin{figure*}[tbp]
\centerline{\includegraphics[bb = 0 0 1854 481, width=16cm]{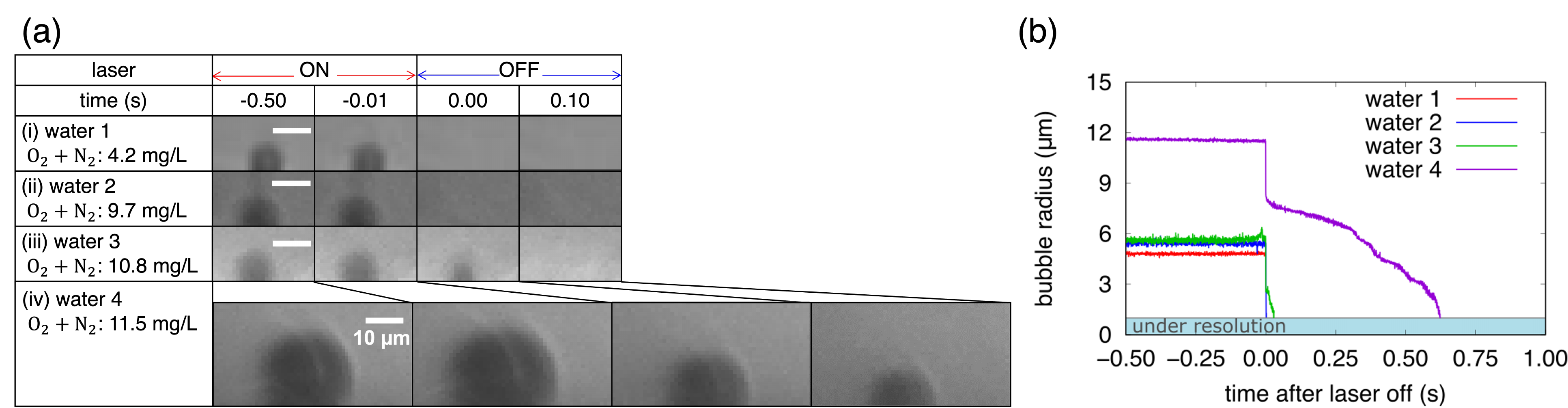}}
\caption{(a) Time evolutions of the bubbles generated in (i) water 1, (ii) water 2, (iii) water 3, (iv) water 4. Time was taken from the moment the laser was turned off. Scale bars: 10 \si{\um}. (b) shows the detailed time dependence of the bubble radius corresponding to the bubbles shown in (a).}
 \label{fig2}
\end{figure*}

Fig. \ref{fig2} shows the typical bubbles observed in water with various non-condensable gas concentration. 
Images were taken at 4000 fps with an exposure time $\sim$0.25 ms.
Note that vapor bubble oscillations with a magnitude of several hundred kilohertz were not captured here. \cite{Namura2020}
This is because the exposure time of the camera was sufficiently long compared with the oscillation period of the bubble.
The oscillating bubble image is averaged over time within the exposure time, resulting in a blurred image.
The bubble size referred to in this study was the size of the bubble, including the blurred image outline of the bubble. If the bubbles oscillated, the measured bubble size was approximately equal to the maximum bubble size during oscillation. 
Under these imaging conditions and the non-condensable gas concentrations listed in Table \ref{tb1}, the bubble size appeared to stabilize 0.01 s after the laser was turned on.
A lower concentration of non-condensable gas dissolved in water tended to lead to a smaller bubble size.
Therefore, the difference in bubble size can be attributed to the difference in the composition of the gases contained within the bubbles, that is, the difference in the amounts of non-condensable gas and water vapor.
To examine the amount of noncondensable gas contained in the generated bubbles, it is best to look at the change in bubble size after stopping laser heating. \cite{Baffou2014,Namura2017}
We observed a temporal change in the bubble diameter before and after the laser was turned off (Fig. \ref{fig2}(a)).
Fig. \ref{fig2}(b) shows the detailed time dependence of the bubble radius corresponding to the bubbles shown in Fig. \ref{fig2}(a).
As mentioned above, the bubble radii were stable during laser illumination at the non-condensable gas concentrations used in this study.
After the laser was turned off, each bubble began to shrink and eventually disappeared. The smaller bubbles shrank faster than the larger bubbles after the laser was turned off. For example, the bubble generated in water 1 takes less than the time resolution of camera 2, which is 0.25 ms, to disappear after turning off the laser. However, it took approximately 0.6 s before the bubble disappeared in water 4. 
The longer lifetime of the bubbles in water 4 was attributed to the non-condensable gases contained in the bubbles, which took time to dissolve back into the water.
To confirm this, quantitative verification of the change in bubble size over time after the laser was turned off was performed.

\begin{figure*}[tbp]
\centerline{\includegraphics[bb = 0 0 850 184, width=16cm]{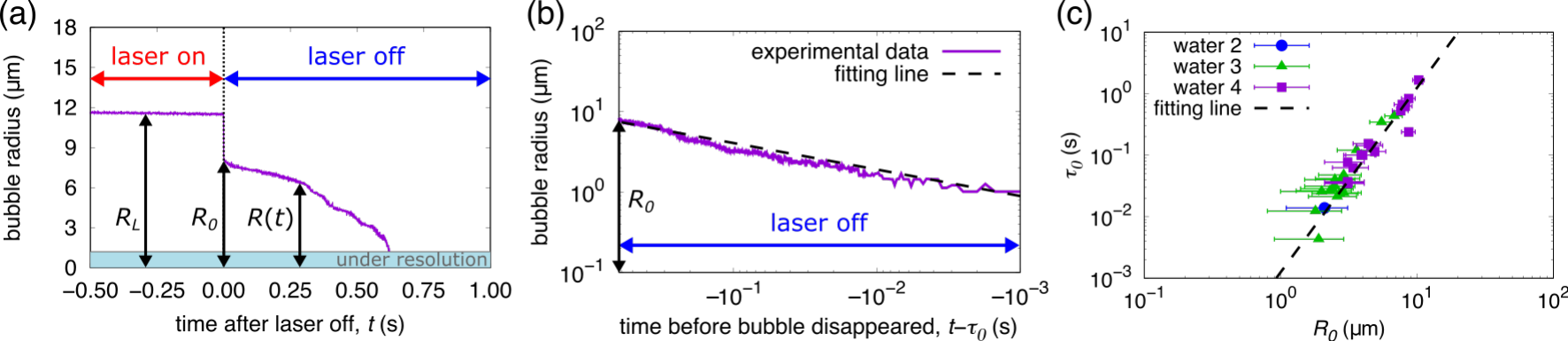}}
\caption{(a) Time dependence of the bubble radius generated in water 4. $R_L$, $R_0$, and $R(t)$ represent the equilibrium radius of the bubble during laser illumination, the radius just after the laser was turned off, and the time-dependent radius after turning off the laser, respectively. (b) The bubble radius as a function of time before the bubbles disappeared, $t-\tau_0$. The black dashed line shows the fit with a slope of 1/3. (c) The lifetime of the bubbles $\tau_0$ plotted as a function of $R_0$ with various non-condensable gas concentrations. The black dashed line shows the fit with a slope of 3.}
 \label{fig4}
\end{figure*}

In Fig. \ref{fig4}(a), we show the representative data of bubble radius changes before and after turning off the laser, where $t = 0$ is the time when the laser is turned off. 
As mentioned previously, the apparent bubble radius during laser illumination $R_L$ was stable. When the laser was turned off, the bubble radius instantly decreased. Some bubbles disappeared when the laser was turned off, and the radii of the other bubbles changed from $R_L$ to $R_0$ when the laser was turned off. 
The immediate decrease in bubble radius can be attributed to the rapid dissipation of heat and the subsequent condensation of water vapor within the bubble.
 This rapid contraction was followed by a relatively slow bubble shrinkage.
The time-dependent bubble radius after the laser was turned off is expressed as $R(t)$. This gradual disappearance of the bubble can be attributed to the diffusion of the non-condensable gas contained in the bubble back into the water. Ljunggren et al. showed that the time dependence of the size of bubbles containing mainly non-condensable gases such as O$_2$ and N$_2$ can be described by the following equations by considering the diffusion of non-condensable gases into water: \cite{Ljunggren1997} 
\begin{equation}
R^3(t) = - \cfrac{6\gamma D A T}{K p_{\infty}} \left(t- \tau_0  \right),
  \label{eq1}
\end{equation}
where $\gamma$, $D$, $A$, $T$, $K$, and $p_{\infty}$ represent the surface tension, diffusion constant of condensable gas, gas constant, temperature, Henry's law constant, and ambient pressure, respectively. The bubble lifetime after the laser is turned off is denoted by $\tau_0$. At $t = \tau_0$, the bubble radius corresponds to zero.
Equation \ref{eq1} indicates that the bubble lifetime is proportional to the third power of the radius.
If the time variation of the bubble radius after the laser is turned off shows a similar trend to this equation, we know that the the bubble with radius $R_0$ is almost filled with non-condensable gas. Then, from $R_0$, the amount of non-condensable gas contained in the bubble generated by laser heating can be determined.
Therefore, we fitted the equation $R(t)=C_1 \left| t-\tau_0 \right| ^{1/3}$ to the temporal change in the radius of the bubble after the laser was turned off, with $C_1$ as the fitting parameter.
In Fig. \ref{fig4}(b), the violet solid line represents the experimental data of the bubble radius, and the black dashed line represents the fitting line.
A clear consistency between the theoretical model and the experimental results was observed. 
This observation indicates that the relatively slow change in the bubble radius after the laser is turned off corresponds to the dissolution of the non-condensable gas back into water.
This also indicates that the amount of non-condensable gas contained in the bubble produced by laser heating can be calculated from $R_0$.
To further confirm that the above discussion can be applied to other bubbles, the relationship between $R_0$ and bubble lifetime $\tau_0$ measured from the number of observed bubbles is shown in Fig. \ref{fig4}(c).
The blue, green, and purple squares represent the results of bubbles generated in water 2, 3, and 4, respectively.
The error bars indicate the camera resolution.
Even in water with the same amount of dissolved gas, the value of $R_0$ varied for each bubble generation.
However, as mentioned above, the lower the amount of dissolved gas, the smaller $R_0$ tends to be. The value of $R_0$ and $\tau_0$ could not be measured from the data of most bubbles generated in waters 1 and 2 because they disappeared immediately when the laser was turned off. 
Therefore, most of the results for waters 1 and 2 are not shown.
The black dashed line represents the relationship $\tau_0=C_2 R_0 ^{3}$ expected from Equation \ref{eq1} fitted to the experimental results, with $C_2$ as the fitting parameter.
The experimental results were consistent with the fitting line. 
This means that under the conditions of this experiment, $R_0$ always represents the amount of non-condensable gas absorbed into the bubble during laser irradiation. 
Then, it is expected that the size of the bubble during laser irradiation is dependent on $R_0$, that is, the amount of non-condensable gas taken in.
Therefore, we compared the bubble sizes before and after the laser was turned off.

\begin{figure}[tbp]
\centerline{\includegraphics[bb = 0 0 360 252, width=7cm]{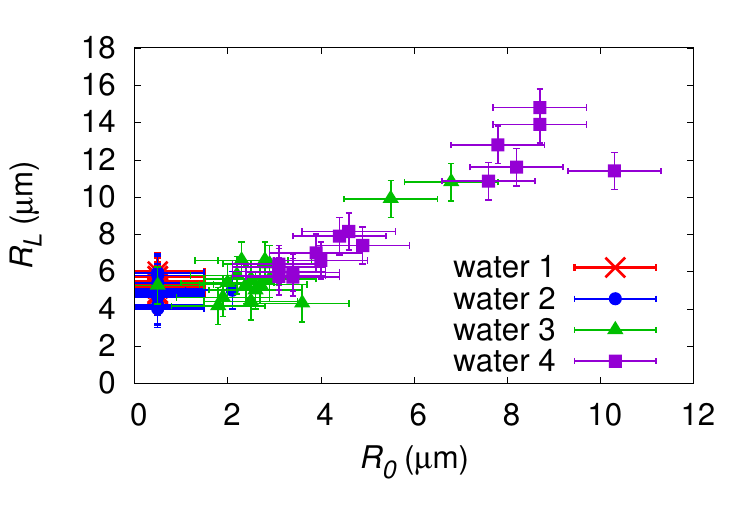}}
\caption{The value of $R_L$ plotted as a function of $R_0$.}
 \label{fig6}
\end{figure}

Figure \ref{fig6} shows the relationship between $R_0$ and $R_L$ measured in waters 1--4.
When $R_0$ was greater than 3 \si{\um}, the value of $R_L$ was approximately proportional to $R_0$. 
This means that the equilibrium radius of the bubble during heating is strongly dependent on $R_0$, that is, the amount of non-condensable gas taken into the bubble. In addition, when $R_0$ was less than 3 \si{\um}, $R_L$ was approximately 5 \si{\um}. It is assumed that bubbles with $R_0$ and $R_L$ smaller than 3 \si{\um} and 5 \si{\um}, respectively, are dominated by water vapor. 
Based on these results, the radius of the bubble varies continuously from 5 \si{\um} to 15 \si{\um}, depending on the amount of non-condensable gas absorbed. In previous studies, we primarily investigated the vapor bubbles formed in well-degassed water, that is, bubbles in water 1. Such bubbles have been found to oscillate at frequencies of sub-megahertz and generate a strong flow around them. On the other hand, it has also been shown that bubbles containing mainly air with diameters of a few tens of micrometers do not oscillate and are not accompanied by a strong flow. 
Then, when the radius of bubbles is gradually increased from 5 \si{\um}m to 15 \si{\um}, to what size do bubbles oscillate?

\begin{figure*}[tbp]
\centerline{\includegraphics[bb = 0 0 502 151, width=12cm]{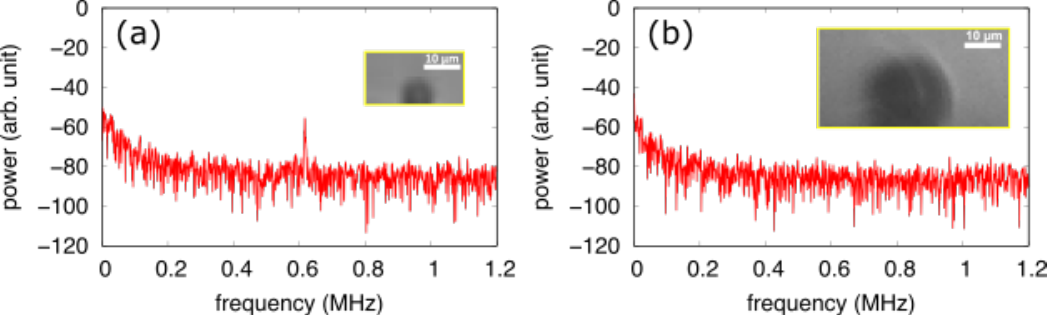}}
\caption{Typical data of the frequency dependence of the temporal variation of the laser light intensity scattered from bubbles with $R_L$ of 11 \si{\um} and 24 \si{\um} generated in (a) water 1 and (b) water 4, respectively. The inset is the image of the corresponding bubble.}
 \label{fig7}
\end{figure*}

Thus, we observed the oscillation of bubbles generated in water with different non-condensable gas concentrations based on the light scattered by the bubbles.  
Figs. \ref{fig7}(a) and \ref{fig7}(b) show the results for bubbles generated in waters 1 and 4 with $R_L$ of 11 \si{\um} and 24 \si{\um}, respectively. 
The data were based on a frequency analysis of the temporal intensity variation of the laser light scattered from the bubbles.
Consistent with a recent study \cite{Namura2020} the bubble containing mainly non-condensable gases showed no peak (Fig. \ref{fig7}(b)), while the bubble containing mainly water vapor shows a clear peak at 620 kHz (Fig. \ref{fig7}(a)). 
This peak frequency corresponded to the oscillation frequency of the bubble. 
The absence of a peak indicates that the bubble did not oscillate or that the amplitude of the oscillation was very small.
We then measured the oscillation frequency of the bubbles with various radii.

\begin{figure*}[tbp]
\centerline{\includegraphics[bb = 0 0 465 145, width=12cm]{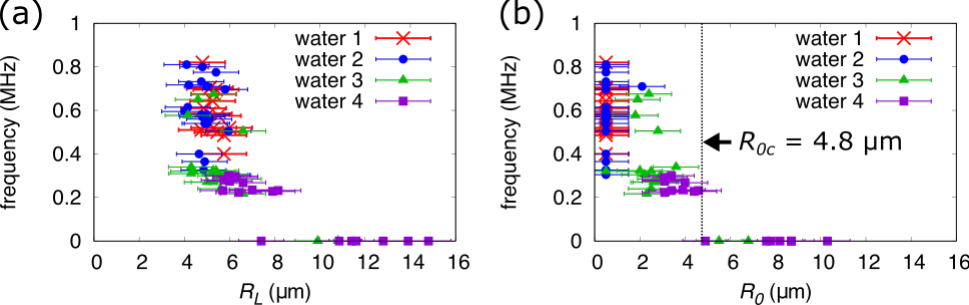}}
\caption{Oscillation frequency of the bubble generated in water with various O$_2$ and N$_2$ concentration plotted as a function of (a) $R_L$ and (b) $R_0$. The threshold of $R_0$ between the oscillating and the non-oscillating bubbles, $R_{0c}$, is 4.8 \si{\um}. }
 \label{fig8}
\end{figure*}

In Fig. \ref{fig8}(a), we plot oscillation frequency of the bubble generated in water with various concentration of non-condensable gas as a function of $R_L$.
The error bars indicate the camera resolution.
 In waters 1 and 2, most of the bubbles show a relatively small $R_L$ ranging between 4 \si{\um} and 6 \si{\um} and a high oscillation frequency ranging between 300 kHz and 800 kHz. However, in waters 3 and 4, many of the bubbles show a relatively large $R_L$ ranging between 6 \si{\um} and 8 \si{\um}, and a low oscillation frequency ranging between 200 kHz and 300 kHz. 
   However, no bubbles oscillating below 200 kHz were observed.
Some of the bubbles generated in waters 3 and 4 had $R_L$ values greater than 7 \si{\um}, and no oscillations could be confirmed.
These results indicate that there is a threshold for bubble oscillation at approximately $R_L$ of 7 \si{\um} to 8 \si{\um}.
The same comparison shown in Fig. \ref{fig8}(a) is made as a function of $R_0$ in Fig. \ref{fig8}(b).
For cases where the bubble radius decreased immediately after the laser was turned off and $R_0$ could not be measured, it was plotted as $R_0 = 0.5\pm0.5$ \si{\um}, indicating that it was below the resolution of the camera.
An approximate correlation exists between the oscillation frequency and $R_0$; the oscillation frequency increases as $R_0$ decreases. 
No bubble oscillation was observed when $R_0$ was greater than $R_{0c} = 4.8$  \si{\um}. 
This result indicated that the bubble did not oscillate when the bubble equilibrium radius during laser heating exceeded a certain size.
The equilibrium radius during laser heating is determined by the amount of non-condensable gas, indicated by $R_0$, in addition to the water vapor present in the bubble.
Taking Fig. \ref{fig8}(a) into account, we can say that the bubble stops oscillating when the equilibrium radius of the bubble is larger than 7--8 \si{\um}.
Assuming that the bubble oscillation is due to repeated evaporation and condensation, for the bubbles to not oscillate, the gas-liquid interface of the bubbles should not be heated above the boiling point when the equilibrium radius is reached.
When the gas-liquid surface of a bubble contacts a heating spot above the boiling point, the water near the bubble surface evaporates. Subsequently, the bubble expands, and the bubble surface moves away from the heating spot. When the heating spot was covered by the gas phase, heat dissipation from the heating spot to the water was insulated. Therefore, the water vapor condenses to water, the bubbles shrink, and the gas-liquid surface reaches the heating spot again. This process repeated during the continuous heating, resulting in periodic bubble oscillations.
However, if a certain amount of non-condensable gas is taken into the bubble and the gas-liquid interface is not heated above the boiling point when the bubble is at its equilibrium radius, no such oscillations occur.
Measuring the temperature distribution around microbubbles is not easy, but future measurements will bring us closer to understanding the oscillation mechanism.
The above results show the effect of the amount of gas dissolved in water on the gas composition in the bubble and bubble oscillation.
Because water vapor oscillating bubbles are known to generate rapid flows, clarifying the relationship between bubble oscillation and the velocity of the water flow generated near the bubbles should be considered in future studies.

\section{CONCLUSION}
In summary, we investigated the dependence of the size and oscillation frequency of a microbubble on the concentration of a non-condensable gas dissolved in water by focusing a CW laser on a $\beta$-FeSi$_2$ thin film. We found that a lower concentration of non-condensable gas dissolved in water led to a smaller bubble size and a higher oscillation frequency. Furthermore, we demonstrated that the amount of non-condensable gas absorbed into the bubble strongly affected the oscillation mechanism of the bubble. 
There was a clear threshold for the amount of non-condensable gas taken up by a bubble that determined whether the bubble would oscillate. 
The bubble oscillates when the amount of non-condensable gas absorbed into the bubble is sufficiently small for the bubble surface to be directly heated on the heating spot. 
These results will be helpful for developing novel cooling systems using microfluidics.

\begin{acknowledgments}
This study was supported by the JST FOREST Program (Grant Number JPMJFR203N, Japan) and JSPS KAKENHI Grant Nos. 19K21932 and 21H01784. 
It was also financially supported by a collaborative research project between Kyoto University and Mitsubishi Electric Corporation on Evolutionary Mechanical System Technology. 
\end{acknowledgments}


\bibliography{2023hiroshige}

\end{document}